\documentclass[aps,showpacs,showkeys,superscriptaddress]{revtex4}

\usepackage{graphicx}
\usepackage{makeidx}
\usepackage[]{hyperref}
\usepackage{amsmath}

\usepackage{amssymb,epsfig,graphics}
\usepackage{amscd}
\usepackage{verbatim}

\usepackage{amsmath}
\usepackage{amsthm}
\usepackage{amsfonts}

\begin{document}
\title{Electronic structure of disordered graphene with Green's function approach}

\author{J. Smotlacha}\email{smota@centrum.cz}
\affiliation{Faculty of Nuclear Sciences and Physical Engineering, Czech Technical University, Brehova 7, 110 00 Prague,
Czech Republic}
\affiliation{Bogoliubov Laboratory of Theoretical Physics, Joint
Institute for Nuclear Research, 141980 Dubna, Moscow region, Russia}

\author{R. Pincak}\email{pincak@saske.sk}
\affiliation{Bogoliubov Laboratory of Theoretical Physics, Joint
Institute for Nuclear Research, 141980 Dubna, Moscow region, Russia}
\affiliation{Institute of Experimental Physics, Slovak Academy of Sciences,
Watsonova 47,043 53 Kosice, Slovak Republic}

\author{M. Pudlak}\email{pudlak@saske.sk}
\affiliation{Institute of Experimental Physics, Slovak Academy of
Sciences, Watsonova 47,043 53 Kosice, Slovak Republic}

\date{\today}



\def\wu{\widetilde{u}}
\def\wv{\widetilde{v}}

\begin{abstract}
The Green functions play a big role in the calculation of the local density of states of the carbon nanostructures. We investigate their nature for the variously oriented and disclinated graphene-like surface. Next, we investigate the case of a small perturbation generated by two heptagonal defects and from the character of the local density of states in the border sites of these defects we derive their minimal and maximal distance on the perturbed cylindrical surface. For this purpose, we transform the given surface into a chain using the Haydock recursion method. We will suppose only the nearest-neighbor interactions between the atom orbitals, in other words, the calculations suppose the short-range potential.
\end{abstract}

\maketitle

\section{Introduction}\

The local density density of states ($LDoS$) is one of the most important characteristics describing the electronic properties of the carbon nanostructures. Different methods were used for its calculation: The first exploits the form of the electronic spectra \cite{wakabayashi1}, the second deals with the gauge-theory model and the Dirac equation \cite{osipov, vanhove}, the third works with the Green function which can be calculated using different methods.

Possible procedures of the calculation of the Green function can be seen in \cite{disordered} for the case of the presence of the impurity potentials and in \cite{vozmed} for the case of the smooth ripples present in the graphene structure. In this paper, we use the Haydock recursion method \cite{wakabayashi2, tamura} for this purpose. It will be applied for the calculation of the $LDoS$ of different forms of the carbon nanocylinders and other kinds of nanostructured surfaces which arise by adding $2$ heptagonal defects.

First, we describe the Haydock recursion method and the procedure of the calculation of the Green function. Then we apply this method on the calculation of the Green function and related quantities in the edge sites of the carbon nanocylinder and of the graphene nanoribbon perturbed by two heptagonal defects. Then we investigate the changes of the $LDoS$ for the changing distance of the defects, calculate the zero modes and after that we estimate the minimal and maximal distance of the defects on the perturbed surface of the nanocylinder.

\section{Haydock recursion method}\

The $LDoS$ can be defined as
\begin{equation}\label{LDOS}LDoS(E)=\lim\limits_{\delta\rightarrow +0}\frac{1}{\pi}{\rm Im}G_{00}(E-i\delta),\end{equation}
where $G_{00}(E)$ is the Green function. It can be calculated using the recursion procedure which transforms an arbitrary surface into $1-$dimensional chain. This procedure is called the Haydock recursion method \cite{haydock}. It divides the positions of the investigated surface into the groups of sites, each of them represents the site in the $1-$dimensional chain. The investigated site we label by the number $1$ and it lies in the beginning of the chain. The site $2$ in the chain corresponds to the nearest neighbors, the site $3$ corresponds to the next-nearest neighbors etc. For this purpose, we suppose only short-distance interactions. On this base, we can write the action of the Hamiltonian on the $n-$th site in the form
\begin{equation}\label{1}H|n\rangle=a_n|n\rangle+b_{n-1}|n-1\rangle+|n+1\rangle,\end{equation}
where $a_0=b_0=b_{-1}=0$. Then, from the knowledge of the state $|1\rangle$, which corresponds to the usual state of the carbon atom, we can recursively compute the coefficients $a_n, b_n$ corresponding to the particular sites of the chain using
\begin{equation}\label{2}|n+1\rangle=(H-a_n)|n\rangle-b_{n-1}|n-1\rangle.\end{equation}
The maximal value of $n$ which is $n_{max}$ determines the recursion depth. It is given by the size of the concrete surface, but in the case of infinitely large graphene, nanocone etc., it is up to our choice and it provides the rate of precision.  Then we define $G_{00}(E)$ as \cite{tamura}
\begin{equation}\label{G1}G_{00}(E)=\frac{1}{E-a_1-b_1g_1(E)},\end{equation}
where
\begin{equation}g_1(E)=\frac{1}{E-a_2-b_2g_2(E)},\end{equation}
\[\vdots\]
\begin{equation}g_{n-1}(E)=\frac{1}{E-a_n-b_ng_n(E)},\end{equation}
\begin{equation}\label{Gn}g_{n_{max}-1}(E)=\frac{E-a_{n_{max}}}{2b_{n_{max}}}\left(1-\sqrt{1-\frac{4b_{n_{max}}}{(E-a_{n_{max}})^2}}
\right).
\end{equation}
It can be found from (\ref{2}) that for $1\leq n\leq n_{max}$,
\begin{equation}\label{ab}a_n=\frac{\langle n|H|n\rangle}{\langle n|n\rangle},\hspace{1cm}b_n=\frac{\langle n|H|n+1\rangle}{\langle n|n\rangle}=\frac{\langle n+1|n+1\rangle}{\langle n|n\rangle}.
\end{equation}
To calculate $\langle n|H|n\rangle$, knowledge of the expressions $\langle n|H^2|n\rangle$, $\langle n|H^3|n\rangle$ will be needed. They have the form
\begin{equation}\label{sum}\langle n|H^r|n\rangle=\sum\limits_{n_1,n_2,\ldots,n_{r-1}}\langle n|H|n_1\rangle\langle n_1|H|n_2\rangle\ldots\langle n_{r-1}|H|n\rangle,\end{equation}
where the sum goes over all nonzero possibilities, in other words, each term of the sum is formed by a product of the cycles containing the atom from the $n$-th site, where the total number of the atoms in the term is $2$ or $3$, respectively. The form of (\ref{sum}) is characteristic for each kind of the nanostructures.

\section{LDoS of nanocylinder}\

The carbon nanocylinder is a nanotube without a cap and with a finite length. It arises by rolling up a graphene sheet. Then, the atomic structure of the molecular surface depends on the orientation of this graphene sheet. Similarly as for the nanotubes, we distinguish three forms of the nanocylinders: armchair ($ac$), zig-zag ($zz$) and achiral \cite{saito}.

\begin{figure}
{\includegraphics[width=150mm]{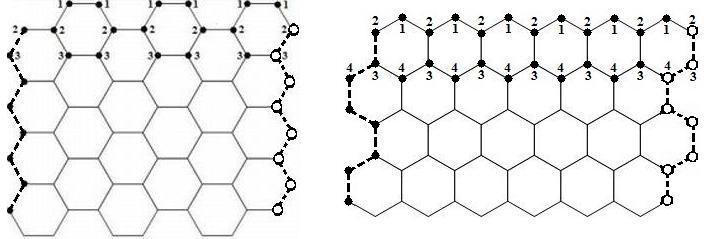}}\caption{Surface of two forms of the carbon nanocylinders: armchair (left) and zig-zag (right); the labeling of the sites corresponds to the technique described in the chapter II; there are equivalent sites in each line parallel with the edge and that is why we label each line by the same number; the dashed lines consisting of sites denoted by black or white color are identical on the real surface.}\label{fg0}
\end{figure}

In this paper, we will be concerned with the armchair and the zig-zag form. In Fig. \ref{fg0}, we see the surface of these two forms together with the labeling of the sites in accordance with the technique described in the previous chapter. The armchair form should be always metallic, the zig-zag form is mostly semimetallic and rarely metallic. The evidence of the metallicity is given by the peak in the $LDoS$ for the Fermi level \cite{wakabayashi1}.

\begin{figure}
{\includegraphics[width=120mm]{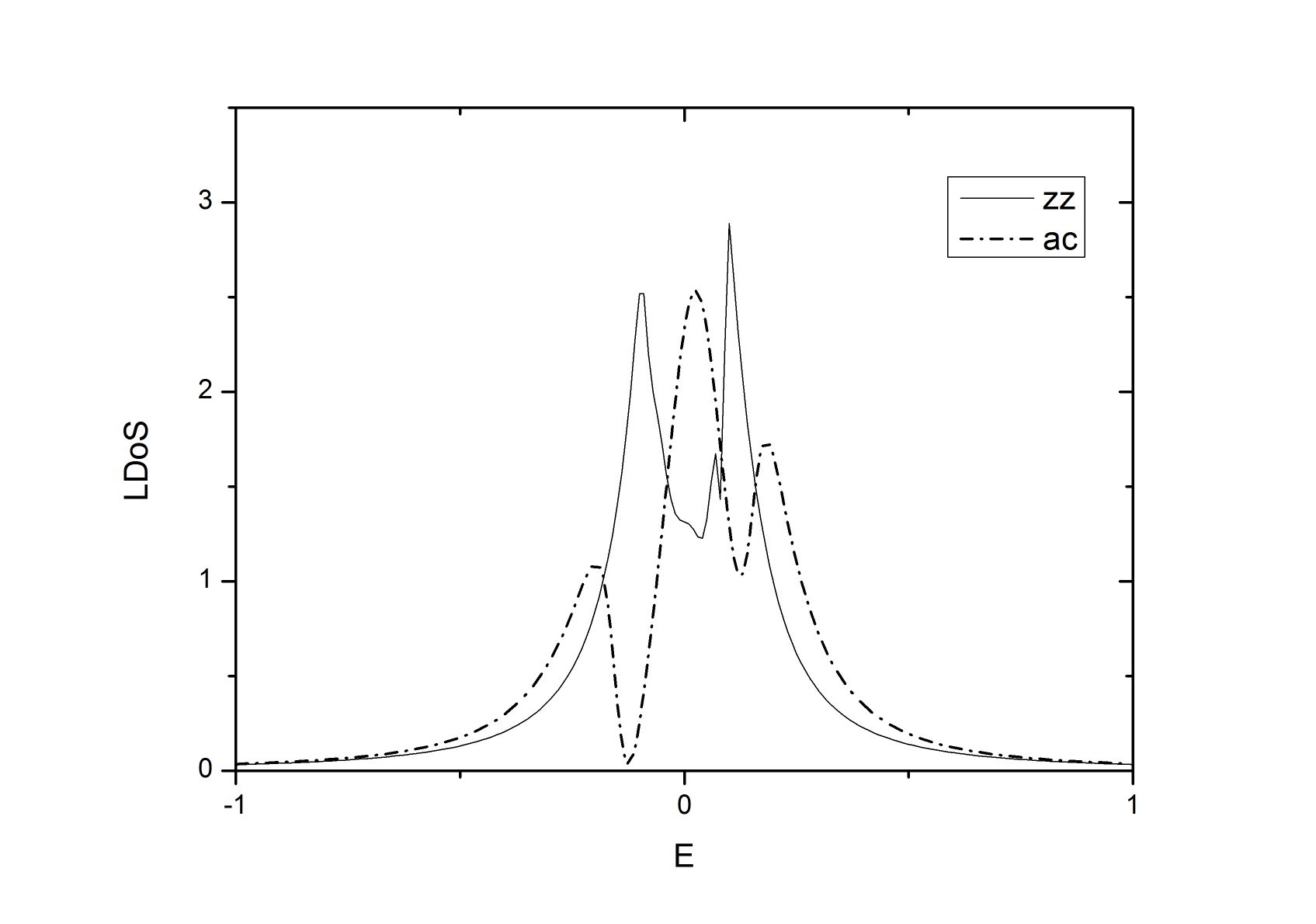}}\caption{$LDoS$ for armchair and zig-zag cylinder; longitudinal number of atoms: 12, circumferential number of atoms: 10 for armchair, 20 for zig-zag; here, $\delta=0.1$.}\label{fg1}
\end{figure}

To apply the Haydock recursion method, we have to choose the recursion depth $n_{max}$, which closely corresponds to the length of the nanocylinder. The procedure of the calculation will differ in the form of the expressions $\langle n|H^2|n\rangle, \langle n|H^3|n\rangle$ in (\ref{sum}) included in the resulting expressions for the calculation of the coefficients $a_n, b_n$. The $LDoS$ for different forms of the nanocylinder is shown in Fig. \ref{fg1} together with the chosen values of the circumferential and the longitudinal number of atoms. The chosen value of the parameter $\delta$ in (\ref{LDOS}) is $0.1$.

\subsection{The case of perturbation}\

\begin{figure}
{\includegraphics[width=150mm]{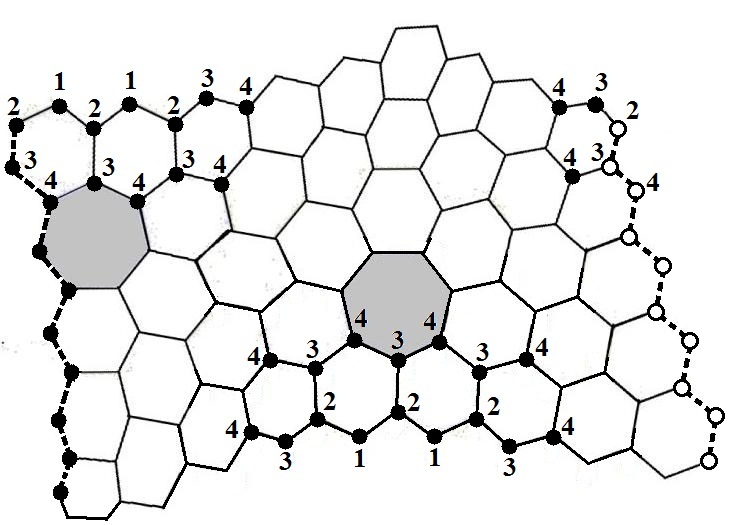}}\caption{Surface of the nanocylinder with a small perturbation; due to the mirror symmetry, we have pairs of equivalent sites in each line parallel with the edge, but there is not any line composed of equivalent sites only; so, we distinguish only the sites which are neighboring, next-neighboring etc. with the site 1 for which the $LDoS$ we calculate; the whole number of the sites in the chain is 9; in the case of the semi closed, nanocylindrical structure, the dashed lines consisting of sites denoted by black or white color are identical on the real surface.}\label{fg3}
\end{figure}

\begin{figure}
{\includegraphics[width=120mm]{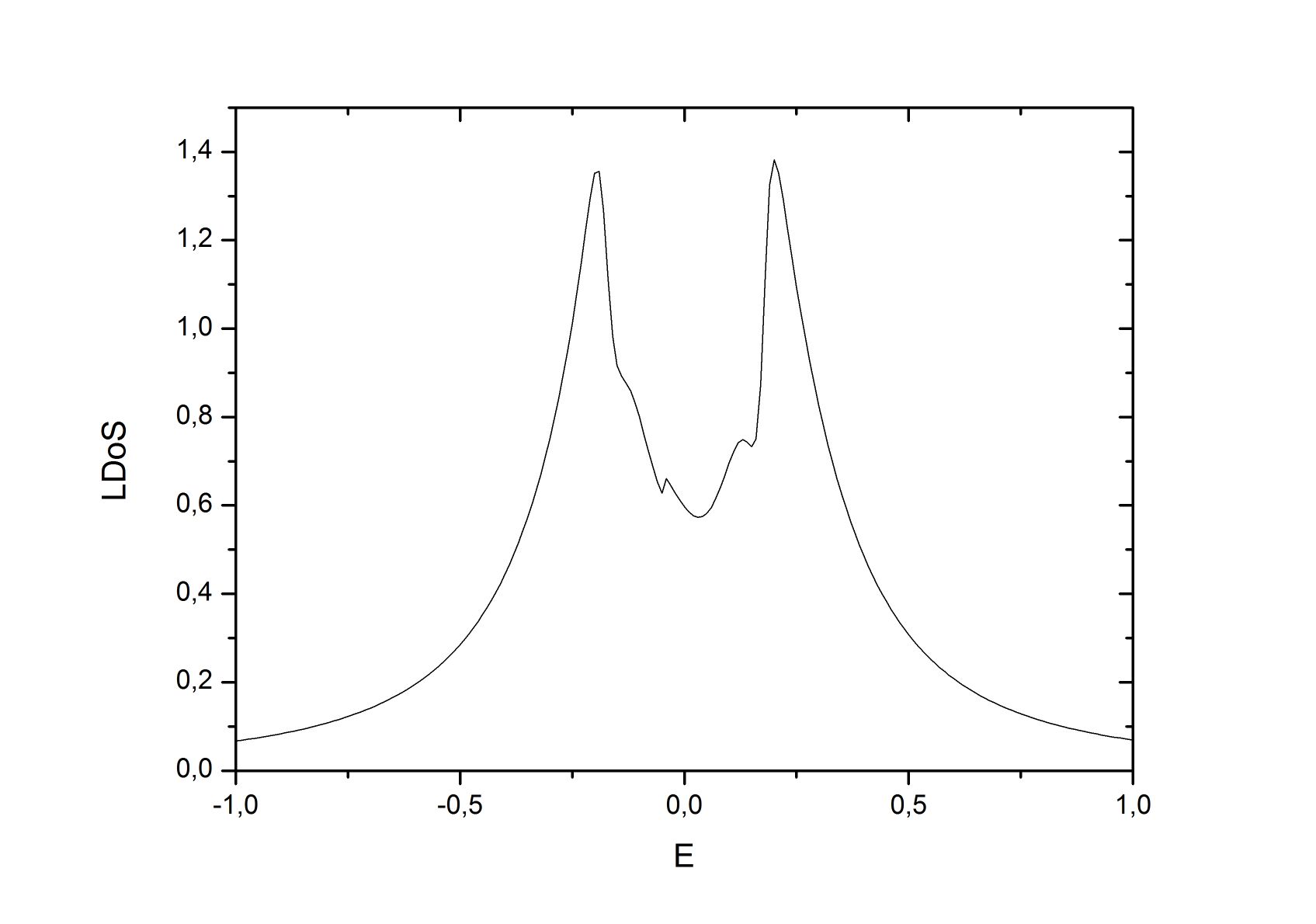}}\caption{$LDoS$ of the perturbed cylinder with surface depicted in Fig. \ref{fg3}; here, $\delta=0.2$.}\label{fg4}
\end{figure}

Let us investigate the $LDoS$ in the edge sites of a perturbed graphene nanoribbon of the sizes which have the same values as the above mentioned cylindrical surface (see Fig. \ref{fg3}). The perturbation is created by two heptagonal defects. From the sketch it is evident that we can't distinct the armchair and the zig-zag edges for this kind of perturbation. Because the structure of the surface is different from the previous case (Fig. \ref{fg0}), the placement and labeling of the equivalent sites is changed. For the chosen edge sites, the result is presented in Fig. \ref{fg4}. In this case, the chosen value of the parameter $\delta$ is $0.2$.

\begin{figure}
{\includegraphics[width=150mm]{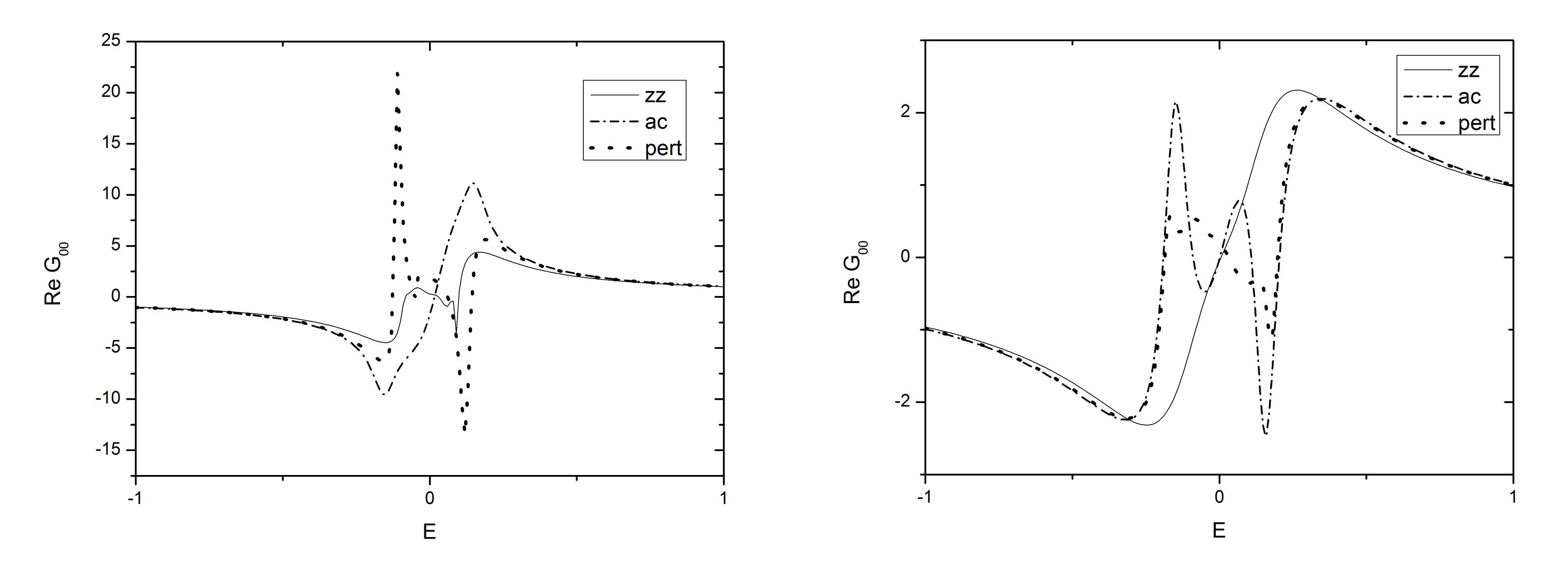}}\caption{Real part of the Green function for armchair, zig-zag and perturbed cylinder with different values of $\delta$: $\delta=0.1$ (left) and $\delta=0.2$ (right)}\label{fg5}
\end{figure}

It is also interesting to compare the nature of the real part of the Green function in all of the investigated cases. The corresponding plots we see in Fig. \ref{fg5}. It strongly depends on the chosen value of $\delta$ which gives the precision of the calculations: the lower $\delta$, the more precise results we get.

\begin{figure}
{\includegraphics[width=150mm]{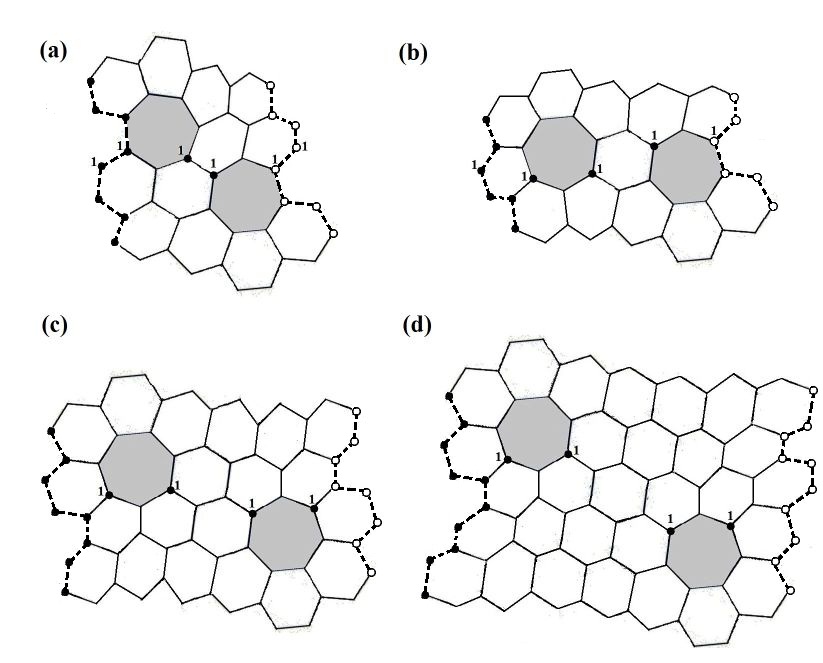}}\caption{Perturbed nanostructured surfaces with different distances of the defects. We calculate the $LDoS$ for the denoted sites; in the case of the semi closed, nanocylindrical structure, the dashed lines consisting of sites denoted by black or white color are identical on the real surface.}\label{fg2hept}
\end{figure}

The limiting sizes of the nanocylinder disclinated by the investigated kind of perturbation will be derived now. For this purpose, we investigate the $LDoS$ in the sites of the defects denoted by number $1$ in the disclinated surfaces depicted in Fig. \ref{fg2hept} and we compare the results with the results presented in \cite{wakabayashi1}, where the $LDoS$ for the simple graphene was presented. In the calculations we suppose that the left and the right parts of the particular surfaces (Fig. \ref{fg2hept}) are joined together.

\begin{figure}
{\includegraphics[width=150mm]{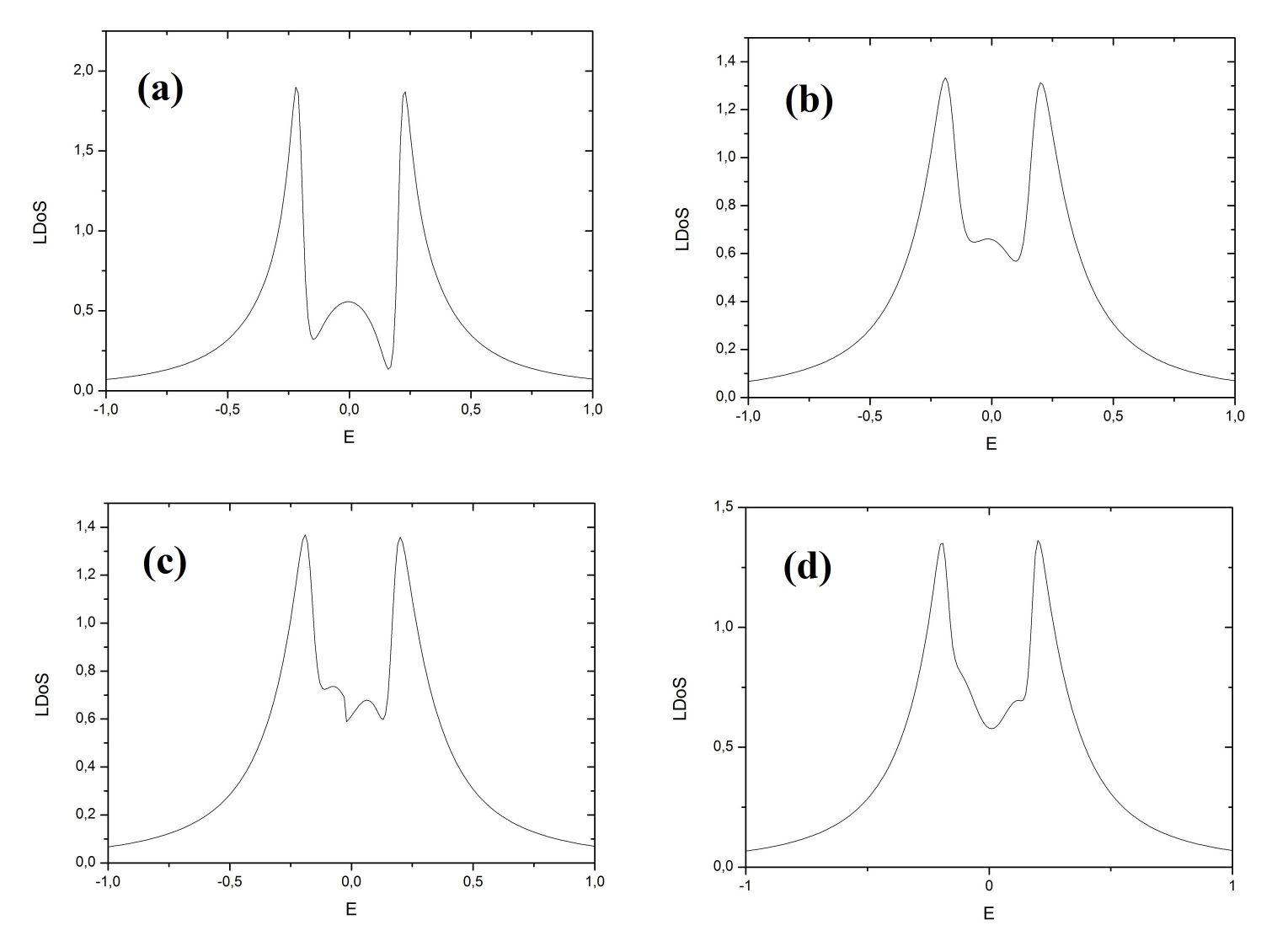}}\caption{$LDoS$ for the particular cases of the perturbed cylindrical surfaces. The notation (a)-(d) corresponds to Fig. \ref{fg2hept}. The value of the parameter $\delta$ is $0.2$.}\label{LDoS2hept}
\end{figure}

In Fig. \ref{fg2hept}, we define the distance of the defects as the distance of the investigated sites lying in different defects. The unit of distance will be given by the distance of the neighboring sites. Using the Haydock recursion scheme, we get the plots of the $LDoS$ outlined in Fig. \ref{LDoS2hept}. Now, it is important to stress that the defects are placed in the middle parts of the nanocylinders. The acquired results should be similar to the $LDoS$ of simple graphene \cite{wakabayashi1}. Then, we suppose the presence of the local minimum for the Fermi level in the corresponding plot.

Let us look through the plots of the $LDoS$ in Fig. \ref{LDoS2hept}. From these plots we see that the growing distance of the defects causes decrease of the $LDoS$ for the Fermi energy and violation of the peak. The case (d) in Fig. \ref{LDoS2hept} corresponds to the expected shape of the $LDoS$ \cite{wakabayashi1}. From this follows an important conclusion that the surface (d) in Fig. \ref{fg2hept} corresponds to the minimal necessary size of the cylindrical surface perturbed by $2$ heptagonal defects and so, the minimal distance between the defects on the perturbed cylindrical surface is $4$ times the distance of the nearest neighbor atoms.

\begin{figure}
{\includegraphics[width=150mm]{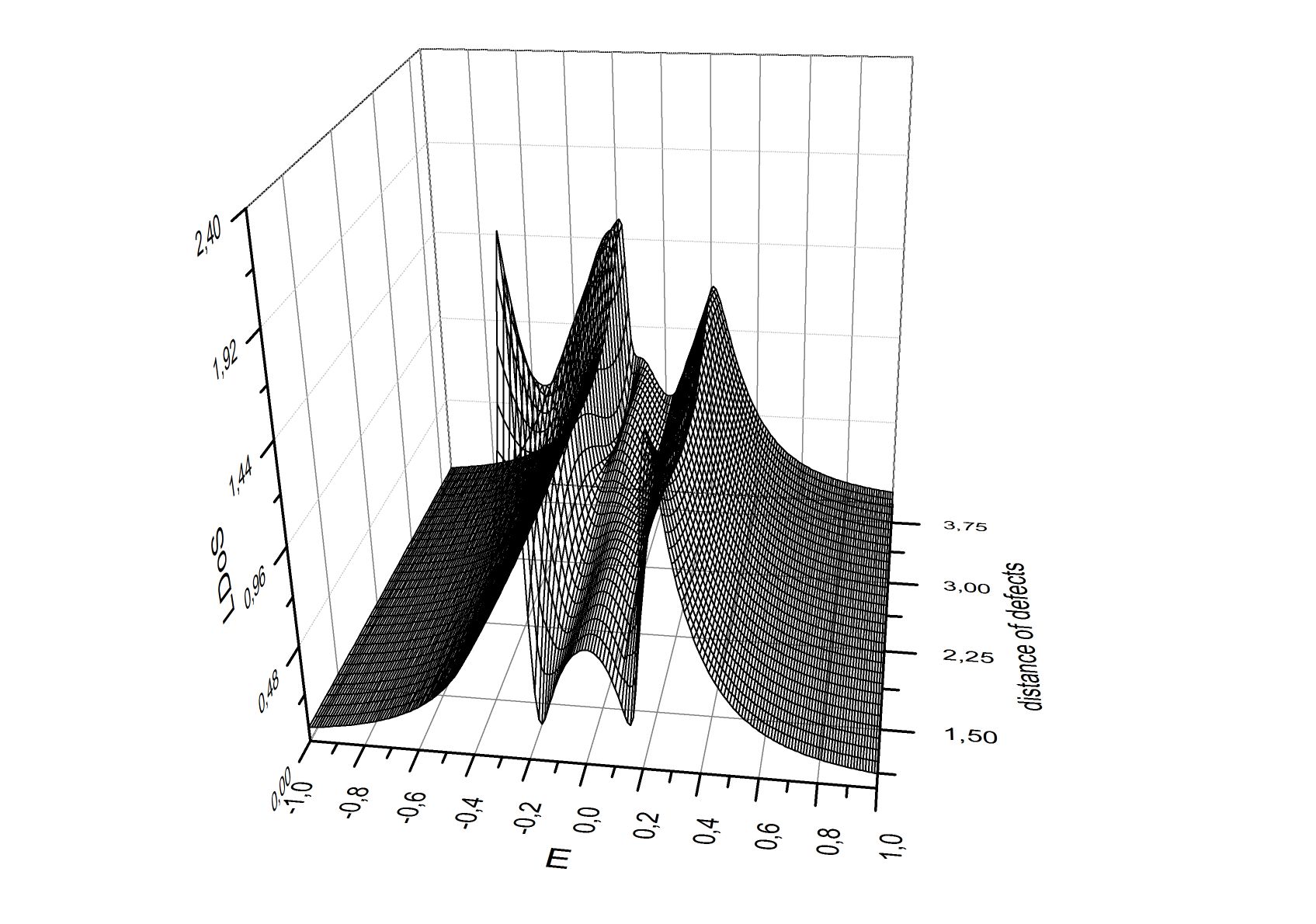}}\caption{Approximation of the dependence of the $LDoS$ on the energy and on the distance of the defects (with the unit given by the distance of the nearest neighbor atoms). Here we see the evidence of the decrease of the $LDoS$ with the growing distance of the defects.}\label{3DLDoS}
\end{figure}

If we do an approximation of the dependence of the $LDoS$ on the distance of defects, we get the 3D plot in Fig. \ref{3DLDoS}. From this plot follows the decrease of the $LDoS$ with the growing distance of defects. We can estimate from the character of this decrease that the $LDoS$ violates for the distance of defects which corresponds to $8$ chosen units of length, i.e. for the surface which is twice longer than the surface (d) in Fig. \ref{fg2hept}. We can suppose that this case corresponds to the maximal permissible distance of the defects in the perturbed cylinder.

\subsection{Zero modes}\

\begin{figure}
{\includegraphics[width=120mm]{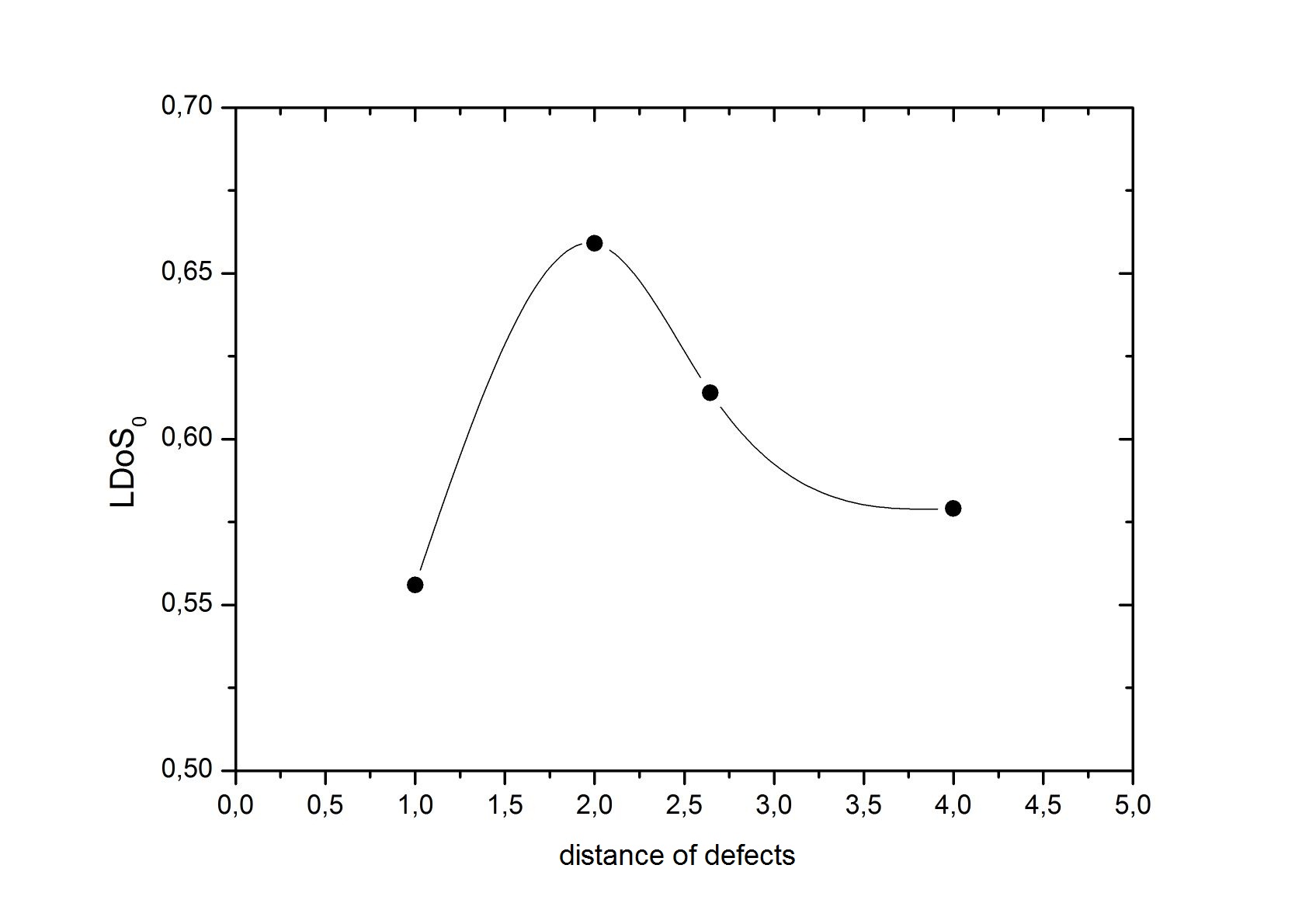}}\caption{$LDoS$ of the zero modes ($LDoS_0$) depending on the distance of the defects which is understood as the distance of the investigated sites lying in different defects (see Fig. \ref{fg2hept}). Here the unit distance is given by the distance of the nearest neighbor atoms.}\label{nulmods}
\end{figure}

We denote the $LDoS$ for zero energy as $LDoS_0$. From the outlined plots, it is possible to calculate the $LDoS_0$ for all of the investigated cases. Next, for the surfaces in Fig. \ref{fg2hept}, we can find the dependence of $LDoS_0$ on the distance of defects. The result we see in Fig. \ref{nulmods}.

We can also compare $LDoS_0$ in the edge sites corresponding to the case of armchair, zig-zag and perturbation. The concrete values for $\delta=0.1$ we see in Table I. In accordance with our expectation, the highest value corresponds to armchair which has metallic character. The lowest value corresponds to the perturbation. But on the edge of the perturbed cylinder are not equivalent sites, so this value is changing from site to site. It could be interesting to investigate the zero modes in all of the edge sites of the perturbed cylinder.

\begin{table}

\caption{Zero modes of the $LDoS$ in the edge sites for $\delta=0.1$. $LDoS_0$ is present in the units $(\frac{\hbar v_F}{a})^{-1}A^{-2}$, where $\hbar$ is the Planck constant, $v_F$ is the Fermi velocity (taken as $1$), $a$ is the size of the surface, $A$ is angstrom.}
\begin{tabular*}{0.55\textwidth}{@{\extracolsep{1cm}}c c c c c}
\hline
& & $ac$ & $zz$ & perturbation \\
\hline
& $LDoS_0$ & $2.34$ & $1.31$ & $0.32$ \\
\hline
\end{tabular*}

\end{table}

\section{Conclusion}\

We applied the Haydock recursion method on the calculation of the $LDoS$ of the carbon nanocylinder. We can compare the results presented in Fig. \ref{fg1} with the calculation in \cite{wakabayashi1}, where the form of the electronic spectrum is applied. The results presented in this paper are close to our results. They are also similar to the plots presented in \cite{ryndyk}. In both of these papers as well as in Fig. \ref{fg1}, the difference between the armchair and the zig-zag form is given by the peak for the armchair form at the Fermi level. But in Fig. \ref{fg1}, the peak at the Fermi level should be much closer. The inaccuracy is given by the choice of the values of $\delta$ and of the parameters $a_n, b_n$ in the Haydock recursion method which does not provide a single solution.

Next, we derived that the minimal size of the disclinated cylindrical surface containing $2$ heptagonal defects corresponds to the case (d) in Fig. \ref{fg2hept} and that the maximal size corresponds to the surface which is twice longer. This is also confirmed by the plot of zero modes in Fig. \ref{nulmods}: for the growing distance of the defects, after strong rise in the beginning, the magnitude of the $LDoS_0$ is decreasing.

The model of $2$ defects can be also applied on a simulation of a dipole or a quadrupole present on a defect-free graphene surface: the dipole can be given by a combination of one pentagonal and one heptagonal defect and the quadrupole by two pentagonal and two heptagonal defects. Of course, higher number of defects can give much more possibilities. In the future, the calculations will be focused on these problems.

The plots of the real part of the Green function in Fig. \ref{fg5} are similar to the results for the self-energy in \cite{zhu} and for the real parts in \cite{ryndyk}. It indicates a close connection of the presented results with the case of the disordered graphene. Although we made the calculations on the cylindrical surface, in fact, there was not any indication of the curvature in the procedure, so, the results can be compared with other works dealing with the disclinated graphene.

\end{document}